\documentclass[prl,aps,reprint]{revtex4-1}

\usepackage{amsmath,amssymb,graphicx,color}
\usepackage{soul} 
\begin{document}

\title{Observing half and integer polarization vortices at band degeneracies}

\author{Ang Chen$^{1,3}$}
\thanks{These authors contributed equally.}
\author{Wenzhe Liu$^{1,3}$}
\thanks{These authors contributed equally.}
\author{Yiwen Zhang$^{1,3}$}
\thanks{These authors contributed equally.}
\author{Bo Wang$^{1,3}$}
\author{Fang Guan$^{1,3}$}
\author{Xiaohan Liu$^{1,3}$}
\author{Lei Shi$^{1,3}$}
\email{lshi@fudan.edu.cn}
\author{Ling Lu$^{2}$}
\email{linglu@iphy.ac.cn}
\author{Jian Zi$^{1,3}$}
\email{jzi@fudan.edu.cn}
\affiliation{$^{1}$ Department of Physics, Key Laboratory of Micro- and Nano-Photonic Structures (Ministry of
Education), and State Key Laboratory of Surface Physics, Fudan University, Shanghai 200433, China}
\affiliation{$^{2}$ Institute of Physics, Chinese Academy of Sciences/Beijing National Laboratory for Condensed Matter Physics, Beijing 100190, China}
\affiliation{$^{3}$ Collaborative Innovation Center of Advanced Microstructures, Fudan University, Shanghai 200433, China}


\begin{abstract}
Far-field polarization vortices were recently found on singlet bands in the momentum-space of two-dimensional photonic lattices, also known as the dark states and bound states in continuum. Here, we theoretically proposed and experimentally verified the existence of the polarization vortices at the degenerate points of photonic dispersions, whose vortex cores can be radiative bright states.
Half-charged vortices were generated from the Dirac points of $\pi$ Berry phase and integer-charged vortices were generated from a quadratic degeneracy. Using a home-made polarization-resolved momentum-space imaging spectroscopy, we observed the complete evolution of the splitting from one quadratic point to a pair of Dirac cones by tracking the winding of the polarization vectors and the full spectrum of iso-frequency contours.
\end{abstract}

\maketitle

Quantized vortices, the most fundamental topological excitations~\cite{mermin1979topological,kosterlitz1972long}, have recently been observed in the Brillouin zone (BZ) of two-dimensional periodic photonic~\cite{hsu2013observation} and plasmonic~\cite{zhang2017observation} slabs. Such vortices can exist at multiple momentum points across the entire BZ and their topological charges characterize the winding numbers of far-field polarization vectors in momentum space~\cite{zhen2014topological}. The configuration of vortices, a fascinating topological structure, gives a new perspective view on understanding properties of photonic states, such as the topological nature of bound states in the continuum~\cite{hsu2016bound}. So far, the vortices centered at non-degenerate states are widely discussed, while those centered at degenerate states have not been well investigated yet. Due to the extensive existence of degenerate states in photonic bands~\cite{sakoda2005optical}, researching on those vortices is of fundamental significance. It could complement the general picture of polarization vortices and dig out new physical phenomena.

\begin{figure}[t]
\centering
\includegraphics[scale=1]{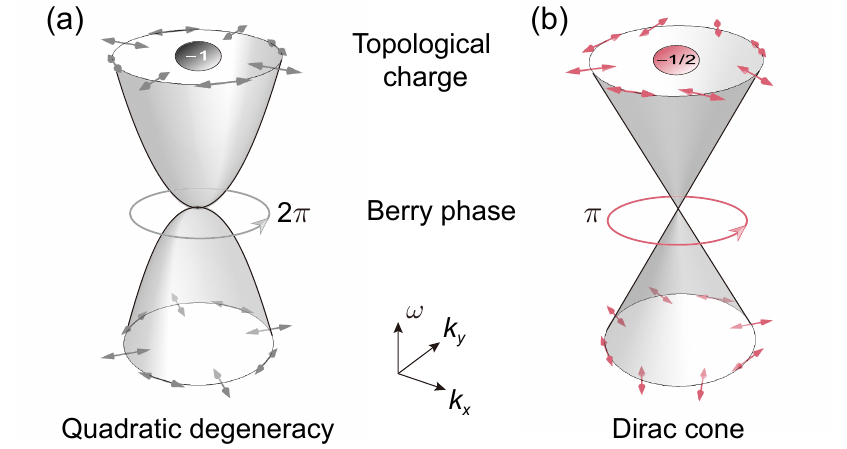}
\caption{Topology of the surface that characterizes the dispersion of (a) quadratic and (b) linear degeneracy points, corresponding to trivial (0 or 2$\pi$) and nontrivial ($\pi$) Berry phase, respectively. The far-field polarizations (denoted by double sided arrows) are shown at the top and the bottom of the surface, with the topological charges shown at the top centers.}
\label{fig:1}
\end{figure}

Degenerate states are classified into two categories according to the dispersions in vicinity: Quadratic degeneracy points and linear degeneracy points (Dirac points), as shown in Fig. \ref{fig:1}. Note that Dirac points induced by accidental degeneracy~\cite{huang2011dirac,sakoda2012proof,zhen2015spawning} are not included here .Those states are topologically different according to their Berry phase which is a phase change acquired over a cycle evolution of Bloch states. Berry phase of quadratic degeneracy points are trivial (0 or 2$\pi$), while that of Dirac points are nontrivial ($\pi$)~\cite{zhang2005experimental,geim2007the,haldane2008possible}. Meanwhile, as the far-field polarization vector represents the zero-order Fourier component of the Bloch state, far-field polarization vortices mentioned above should carry the fingerprint of Berry phase. Therefore, the winding of polarization vectors will be an integer multiple of $\pi$, implying that the polarization vector has to be parallel or anti-parallel to itself after a closed loop centered at the degenerate state. For the quadratic case, it is even, forming a integer vortex that corresponds to trivial Berry phase. On the other hand, the Dirac cones are half vortices corresponding to nontrivial Berry phase.

\begin{figure}[t]
\centering
\includegraphics[scale=1]{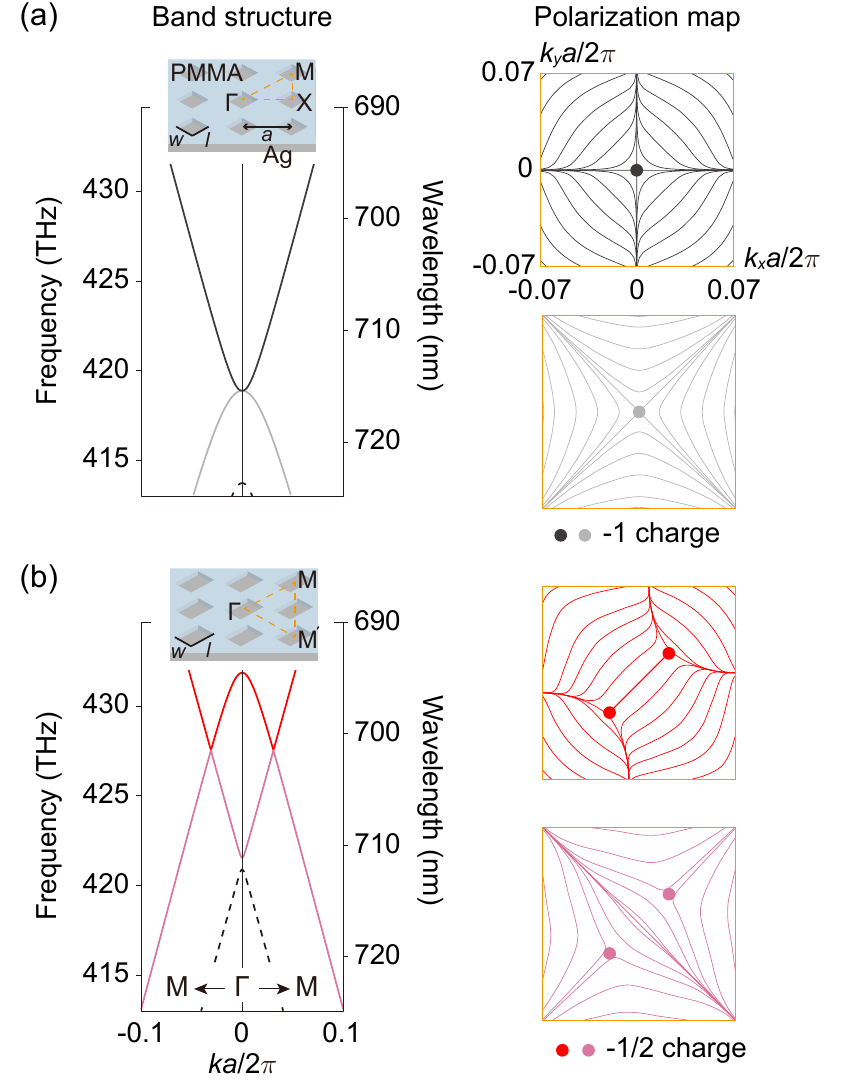}
\caption{Band structures (left panel) and the corresponding polarization maps (right panel) of quadratic and linear degeneracies. The bands of interest are denoted by solid lines, while others by dashed lines. (a) In a plasmonic structure with $C_{4v}$ symmetry, a pair of bands can be degenerate quadratically at $\Gamma$ point. (b) In the $C_{2v}$-symmetry structure, a pair of bands can be degenerate linearly along high-symmetry line (here, $\Gamma$--M direction).}
\label{fig:2}
\end{figure}

In this paper, we theoretically and experimentally demonstrated such vortices, integers at quadratic degeneracy points and half ones at Dirac points in the plasmonic structures. The plasmonic structures studied here  are periodically corrugated thin PMMA layer covered flat silver substrates~\cite{han2006transmission}.
Due to periodicity of the air hole arrays, the surface plasmon polaritons (SPPs) could be excited, forming well-defined band structures.
Because of the specific symmetries ($C_{4v}$ or $C_{6v}$), a pair of bands can be degenerate quadratically at high-symmetry point, such as the center ($\Gamma$ point) of the BZ. This quadratic degeneracy can be further regarded as a pair of linear degeneracies continuable to Dirac points, which can be obtained by symmetry breaking~\cite{chong2008effective}.
Besides, Dirac points can also be found at $C_{3v}$-symmetry point, such as $K$ point in the honeycomb lattice~\cite{haldane2008possible,yu2016surface,dong2016valley}.
Here we took the former approach to observe the evolution from a quadratic degeneracy point splitting into a pair of linear degeneracy points. The corresponding integer vortices splitting into pairs of half vortices is also shown. The sum of those half vortices after symmetry breaking equals to that of integer vortices, approving the law of charge conservation.

\begin{figure}[b]
\centering
\includegraphics[scale=1]{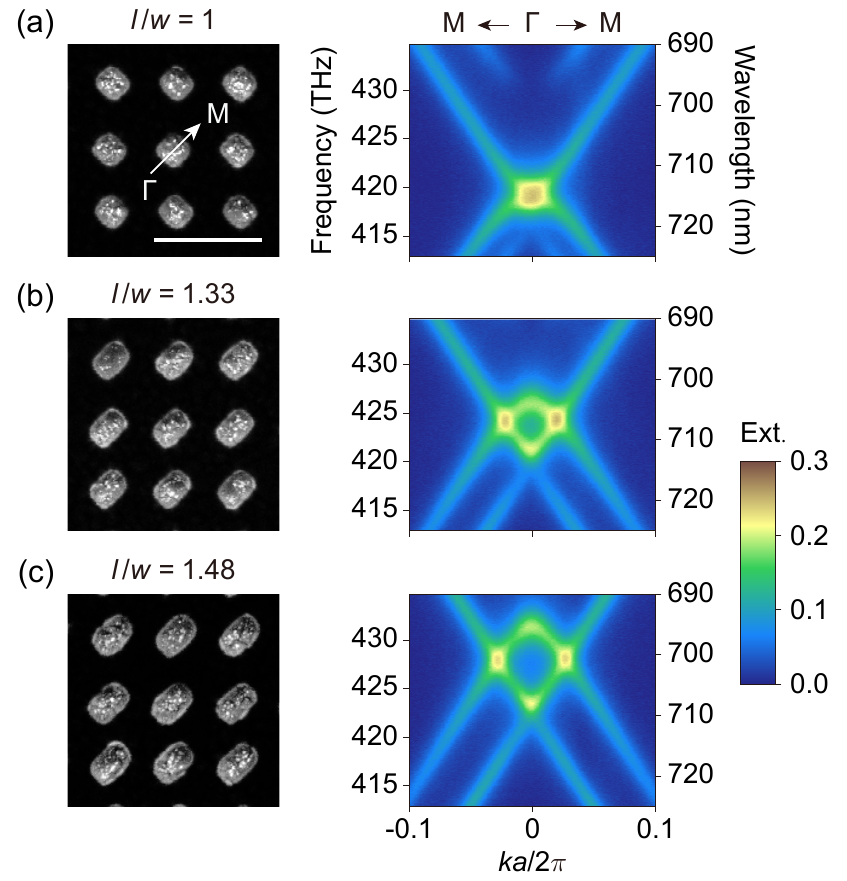}
\caption{The SEM images of the plasmonic structures and experimentally measured band structures (extinction spectra) along $\Gamma$--M direction. From top to bottom, the width ($w$) of the holes stays the same: $w = 290$ nm, while the length $l$ increases: $l = 290,385,430$ nm. The white scale bars: 1 $\mu$m.
(a) $l/w = 1$, corresponds to $C_{4v}$ symmetry. A quadratic degeneracy point appears at $\Gamma$ point. (b) $l/w = 1.33$ and (c) $l/w = 1.48$ correspond to $C_{2v}$ symmetry. As $l/w$ increases, a pair of Dirac points appear along the $\Gamma$--M direction and move towards the M points.}
\label{fig:3}
\end{figure}

\begin{figure*}[t]
\centering
\includegraphics[scale=1]{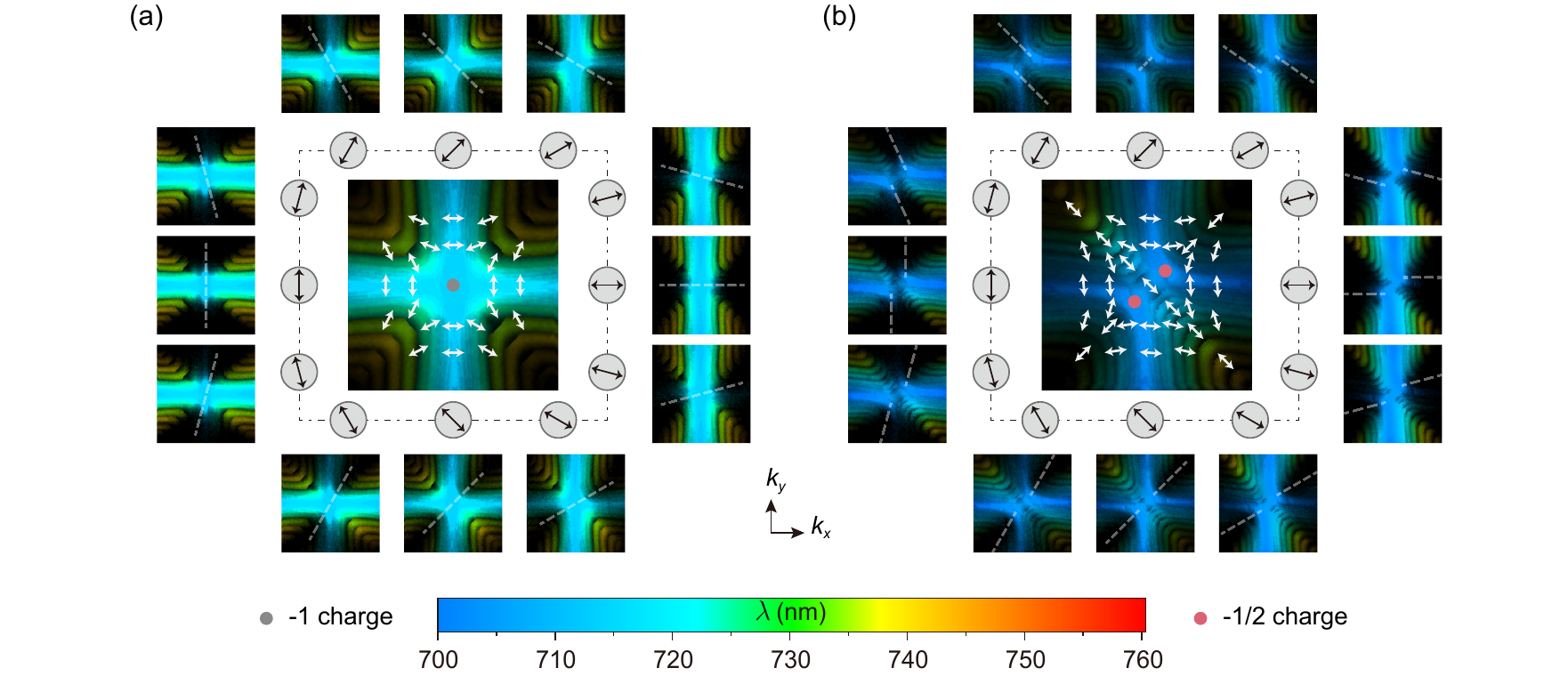}
\caption{Extinction map of the band of interest in the BZ, obtained by summing 12 iso-frequency contours. Central plot: Unpolarized extinction map. Side plots: Polarization resolved extinction maps. The arrows aside indicate the direction of the polarizer. Different colors correspond to different wavelengths, with the colormap shown beside. The centers of integer and half vortices are marked by grey and red dots in the unpolarized plot. Measured polarization vector configurations of those vortices are shown around vortex centers. (a) $C_{4v}$-symmetry system, $l/w = 1$. (b) $C_{2v}$-symmetry system, $l/w = 1.48$.}
\label{fig:4}
\end{figure*}
The band structures along $\Gamma$--M direction and the corresponding polarization vectors of quadratic and linear degeneracies are calculated by using finite element method (COMSOL Multi-physics), as shown in Fig. \ref{fig:2}. The structures studied are illustrated in Fig. \ref{fig:2}(a) and (b). The silver substrates are 200 nm thick, ensuring that there is no light transmitted. The thickness of PMMA is 80 nm, and the period ($a$) of square arrays of rectangular air holes is 600 nm. The length-width ratio $l/w$ ($l, w$ are the length and width of a rectangle hole respectively and the direction of $l$ is along the diagonal lines of the square lattice) of air holes equals 1 and 1.48 in Fig. \ref{fig:2} (a) and (b), respectively. In Fig. \ref{fig:2}(a), when $l/w = 1$, the upper and lower bands are quadratically degenerate at $\Gamma$ point, which is protected by $C_{4v}$ symmetry. The corresponding polarization vectors plotted on a small region of BZ near $\Gamma$ point for both two bands are shown aside. It should be noted that since one cannot distinguish the electric-field direction with its opposite direction due to the time-harmonic oscillations, we represent the polarization vector fields using line segments without arrows, which still allows us to precisely exhibit the vortices. Clearly, the topological charges for those bands at $\Gamma$ point are both $-1$. When the length of the rectangle hole is increased to $l/w = 1.48$ (all other parameters fixed), the symmetry reduces to $C_{2v}$. Consequently, the quadratic degeneracy protected by $C_{4v}$ splits into a pair of Dirac cones along $\Gamma$--M direction, see Fig. \ref{fig:2}(b). From the polarization distributions shown aside, it is found that the topological charge of $-1$ disappears at $\Gamma$ point and a pair of singularities appear along $\Gamma$--M. Those singularities corresponds to Dirac points. The polarization vector is anti-parallel to itself after a closed loop centered at those singularities, indicating $\pi$ winding of the polarization vectors, namely half vortices. The same results are also observed in the case of reducing the $C_{6v}$ symmetry of the hexagonal lattice, shown in Supplemental Material (SM)~\cite{sm}. In general, along with splitting the quadratic degeneracy into pairs of Dirac degeneracy by symmetry breaking, the topological evolution of an integer vortex splitting into a pair of half vortices, consistent with charge conservation, is clearly exhibited, showing their different topological nature.

To experimentally verify above discussions, we performed angle-resolved reflectivity measurements via a momentum-space imaging spectroscopy~\cite{zhang2017observation} on a series of
samples with different $l/w$ fabricated using electron-beam lithography. As shown in Fig. \ref{fig:3}, we plotted the band structures defined by the peaks of the extinction spectra (one minus reflectivity) as a function of wavelength along $\Gamma$--M direction.
When $l/w = 1$, the quadratic degeneracy at $\Gamma$ point is protected by the $C_{4v}$ symmetry of the structure and thus there is only one peak of the extinction spectra in the wavelength range of interest; see Fig. \ref{fig:3}(a).
When $l/w \neq 1$, the $C_{4v}$ symmetry of the structures is broken to $C_{2v}$. At $\Gamma$ point, there are two peaks instead of one peak in the $C_{4v}$ case. Meanwhile, another two peaks emerge along $\Gamma$--M, corresponding to the pair of linear degeneracy points, as shown in Fig. \ref{fig:3}(b) and (c).
When $l/w = 1.33$, the Dirac points whose wavelength is 706 nm appear at wave vector $k_\parallel = \pm 0.023\times2\pi/a$ along $\Gamma$--M.
As $l/w$ becomes larger ($l/w = 1.48$), the Dirac points will move outwards along $\Gamma$--M and appear at $k_\parallel = \pm 0.027\times2\pi/a$, with the wavelength 702 nm. The experimental results along $\Gamma$--M direction agree well with theoretical ones (Fig. \ref{fig:2}).
The plotted experimental data along other directions are provided in SM~\cite{sm}.

To directly observe the vortices at band degeneracies, we measured the polarization-resolved iso-frequency contours and summed them with a range of wavelengths~\cite{zhang2017observation}.
In Fig. \ref{fig:4}, we plotted the summed iso-frequency contours of the lower bands of the $C_{4v}$-symmetry structure ($l/w = 1$) and $C_{2v}$-symmetry structure ($l/w = 1.48$), as shown in Fig. \ref{fig:4}(a) and (b), respectively. The center one is measured under unpolarized illumination, while the ones outside are polarization resolved.
Signal will diminish in the polarization resolved plots as dark areas, due to perpendicular far-field polarization of such states to the polarizer. Here, distribution of polarization vectors forming vortices are plotted in the central figure, and vortex centers are shown. In the case of $C_{4v}$-symmetry structure, as the angle change of polarization vectors is $-2 \pi$ along a closed loop counterclockwise around the vortex center~\cite{zhen2014topological}, this indicates a vortex of $-1$. Meanwhile, we obtain $-1/2$ vortices in the case of $C_{2v}$ symmetry.

These vortices can be viewed as rotating dark patterns along with the polarizer in summed iso-frequency contours, which gives a more vivid picture of winding polarization vectors. The shape, spinning speed and spinning direction of such patterns directly reveal the sign and magnitude of the charges carried by the vortices.
For the $C_{4v}$-symmetry structure, we can see that there is one dark strip spinning around the $\Gamma$ point. Note that the state at the degeneracy point is always excited under any polarization, thus making the dark strip appearing disconnected.
The strip rotates $\pi$ clockwise when the polarizer rotates $\pi$ counterclockwise, indicating the charge of the degeneracy point is $-1$.
As for the $C_{2v}$-symmetry structure, there are two spinning dark radials rather than one strip in the $C_{4v}$ case. And the state at $\Gamma$ point can be either excited or not under different polarizations, which is distinguished from the degenerate state of $C_{4v}$ symmetry.
The radials in $C_{2v}$ system rotate $2\pi$ clockwise when the polarizer rotate $\pi$ counterclockwise, showing the charges here are both $-1/2$.
Evidently when symmetry of the structure is changed from $C_{4v}$ to $C_{2v}$, the $-1$ charge at the quadratic degeneracy point splits into two $-1/2$ charges at the linear degeneracy points, approving charge conservation. Two animations showing those spinning patterns are present in SM~\cite{sm}. For an intuitive understanding, we give a schematic view of the spinning dark patterns for the half vortices; see SM~\cite{sm}.
Last but not least, we emphasize that vortices near both quadratic and linear degeneracies don't necessarily lead to bound states in the continuum, even though the polarizations are ill-defined at those degeneracies.

In summary, we have experimentally observed topological integer and half vortices at band degeneracies of the plasmonic structures in momentum space. Integer vortices could appear near the quadratic degeneracy while half vortices near the linear degeneracy. Via controlling the symmetry of the structure, the evolution from one quadratic degeneracy point to a pair of linear degeneracy points is clearly exhibited, consistent with charge conservation. These unique vortices at band degeneracies enrich the various topological phenomena in photonics~\cite{lu2014topological,xiao2014surface,gao2015topological,leykam2016edge,leykam2016anomalous,he2016photonic,wang2017optical,guo2017three,guo2017topologically} and can accessibly enable half vector-vortex beams~\cite{yao2011orbital,bauer2015observation} or even lasers~\cite{noda2001polarization,iwahashi2011higher,kodigala2017lasing,bahari2017nonreciprocal}.

We thank Yihua Wang, Shaoyu Yin for helpful discussions.
The work was supported by 973 Program (2013CB632701 and 2015CB659400), China National Key Basic Research Program (2016YFA0301100 and 2016YFA0302000) and NSFC (11404064). The research of L. S. was further supported by Science and Technology Commission of Shanghai Municipality (17ZR1442300), Professor of Special Appointment (Eastern Scholar) at Shanghai Institutions of Higher Learning, and the
Recruitment Program of Global Youth Experts (1000 plans).
L.L. was supported by the National key R\&D Program of China under Grant No. 2017YFA0303800, 2016YFA0302400 and supported by NSFC under Project No. 11721404.

\textit{Note}.--A related paper~\cite{zhou2017observation} on topological half charges in the photonic crystal slab came to our attention when the preparation of this work.


%

\end{document}